\documentstyle[preprint,aps,epsf]{revtex}
\tightenlines
\begin{document}
\preprint{
  \parbox{2.0in}{%
       hep-ph/9708464 \\
       MSU-HEP-70826  \\
                      \\}
}
\title{Hard underlying event correction
to inclusive jet cross sections}
\author{Jon Pumplin}
\address{
Department of Physics \& Astronomy \\
Michigan State University \\
East Lansing, MI \\
Internet address:  pumplin@pa.msu.edu
}
\date{\today}
\maketitle
\begin{abstract}
Jets observed in hadron-hadron scattering contain a contribution from
the ``underlying event'' that is produced by spectator interactions
taking place incoherently with the major parton-parton collision, due
to the extended composite structure of the colliding hadrons.  Using a
recent measurement of the double parton interaction rate, we calculate
that the underlying event may be $2$--$3$ times stronger than generally
assumed, as a result of semi-hard perturbative multiple-parton
interactions.  This can have an important influence on the inclusive jet
cross section at moderate values of $E_T$, persisting at the $5-10\%$
level to the largest observable $E_T$.  We show how the underlying event
can be measured accurately using a generalization of the method first
proposed by Marchesini and Webber.
\end{abstract}
\pacs{}
\narrowtext

\section {Introduction}
\label{sec:intro}

The inclusive jet cross section $d\sigma / dE_T$, averaged over a
small range of pseudorapidity, is an important object for study
because it tests perturbative QCD at the highest $Q^2$ scale
currently possible \cite{cdf,d0}.  Beyond its potential for
detecting physics beyond the standard model or confirming QCD, the
jet cross section is even beginning to play a role in the
global data fitting used to measure parton distribution
functions \cite{global,kosower}.  It is therefore important to carefully
consider all systematic effects that influence the interpretation of
the measurement.

Among those effects is the ``underlying event'' generated by spectator
interactions that can occur concurrently with a major parton-parton
collision, due to the extended composite structure of the colliding
hadrons.  Simulations such as the {\footnotesize HERWIG} Monte Carlo
\cite{herwig} include a ``soft'' underlying event that is modeled by
a parametrization of minimum bias
data.\footnote{Minimum bias events provide at best an imperfect model
of the underlying event, since for example they always contain
particles; while the soft underlying event can sometimes be absent,
as shown by the finite survival probability for inelastic events
with large rapidity gaps \cite{gaps}.}
However, there may also be a {\it hard} underlying event:  particles
can be created by incoherent parton-parton interactions at momentum
transfers that are sufficiently large for perturbative QCD to be a
useful approximation, but much smaller than that of the interaction
mainly responsible for a given high $E_T$ jet event.

Attempts have been made in the past to predict underlying event
contributions, including the perturbative part that is the focus of
our attention \cite{sjostrand,gyulassy,butterworth}.  We make a new
estimate here, based on a recent direct measurement \cite{dpi} of
the rate for double parton interactions in hard scattering.  In the
following sections we derive the prediction, show how it can be
independently tested by experiment, and discuss its effect on the
inclusive cross section.

\section {Hard underlying event}
\label{sec:HUE}

To generate typical semi-hard perturbative final states,
{\footnotesize HERWIG 5.8} \cite{herwig} was used to simulate
QCD $2 \to 2$ hard scattering at the Tevatron energy
$\sqrt{s} = 1.8 \, {\rm TeV}$, with the minimum $p_\perp$
parameter set to a fairly small value
{\footnotesize PTMIN}$\,=2.0 \, \rm{GeV}$ that is nevertheless
large enough for comfort with the perturbative calculation.  This
choice yields a cross section $\sigma = 50.6 \, {\rm mb}$, which
is approximately equal to the full inelastic non-diffractive cross
section \cite{nsd}, so we appear to be taking an extreme
point of view in which that cross section is mainly generated by
perturbative (``minijet'') interactions.  The point of view is
actually not so extreme, since including $s$-channel unitarity
effects, e.g., by an eikonal model, would substantially reduce the
cross section on the basis that once one interaction has taken
place, additional interactions do not really add to the inelastic
cross section \cite{schuler,drees}.  The soft underlying event
feature of {\footnotesize HERWIG} was turned
off ({\footnotesize PRSOF}$\, =0$), since our goal is to study
the perturbative part of the underlying event.

If $\sigma_1$ and $\sigma_2$ are cross sections for distinguishable rare
parton-parton interactions, the cross section for both interactions
to occur in the same event can be written as
\begin{eqnarray}
\sigma_{\rm double} =
\frac{\sigma_1 \, \sigma_2}{\sigma_{\rm eff}} \; ,
\label{eq:eq1}
\end{eqnarray}
with the parameter $\sigma_{\rm eff}$ conveying the probability for
double parton interaction.  $\sigma_{\rm eff}$ carries nonperturbative
information beyond the scope of ordinary parton distribution
functions, since it relates to correlations between partons within a
single hadron.  It has recently been measured to be
$\sigma_{\rm eff} = 14.5 \pm 1.7 \, ^{+ \, 1.7} _{- \, 2.3} \,
{\rm mb}$ \cite{dpi}.

For a fixed small value of $\sigma_1$, corresponding to some rare
hard scattering of type 1, Eq.~(\ref{eq:eq1}) gives
$\sigma_2 / \sigma_{\rm eff}$ as the probability for a second rare
scattering of type 2.  If this type 2 scattering is not rare, and
it occurs randomly, the obvious generalization of Eq.~(\ref{eq:eq1})
is a Poisson distribution in the number $n$ of type 2 interactions
that accompany a given type 1 interaction:
\begin{eqnarray}
P_n = \frac{a^n}{n!} e^{-a}
\label{eq:eq2}
\end{eqnarray}
with mean number $a = \sigma_2 / \sigma_{\rm eff}$.

Our model for the hard underlying event therefore consists of
superimposing $n$ of the above minijet events from
{\footnotesize HERWIG}, with a Poisson probability distribution
(\ref{eq:eq2}) of mean $a = 50.6 \, {\rm mb} / 14.5 \, {\rm mb} =
3.49 \,$.  In doing this, we ignore energy-momentum conservation
effects in the sense that the parton momentum distributions
are taken to be the same as in single-interaction events.  This
assumption should be adequate because the minijet interactions don't
require a great deal of energy, and thus come from partons at rather
small $x$.  (More elaborate models \cite{sjostrand,gyulassy}
based on {\footnotesize PYTHIA} \cite{pythia}, and \cite{butterworth}
based on {\footnotesize HERWIG}, have incorporated these
energy-momentum constraints.)  Meanwhile, the Poisson assumption
could actually {\it underestimate} the frequency of 3 or more
interactions, since these may be enhanced by ``hot spots'' --- e.g.,
by strong spatial correlations between the partons in a beam or
target hadron associated with constituent quarks \cite{constituent}.

In this paper, we focus specifically the case of the underlying
event in dijet production.  However, it will be valuable to look also
at $W$, $Z^0$, or lepton pair production, where the hard scattering
makes a color-neutral object, so there is no radiation associated with
final state jets.  These processes also share the advantage that the
hard scattering is initiated by $q \bar q$, which generates less
radiation than $q g$ or $g g$.  Single jet final states such as
$W + 1 \; {\rm jet}$ and $\gamma + 1 \; {\rm jet}$ should also be
studied.

We generate events with jets of transverse energy
$E_T > 100 \, {\rm GeV}$ using {\footnotesize HERWIG}, and look at
the total transverse energy inside a control region cone
$R = \sqrt{(\eta - \eta_0)^2 + (\phi - \phi_0)^2} < 0.7$, which is
the size typically used to define jets in the
pseudorapidity--azimuthal angle ``Lego'' plane.  The underlying event
is by definition uncorrelated with the jets, so to avoid the majority
of the jet $E_T$, the control cone is centered a distance $R = 2.0$
away from both jet axes, in a manner to be described fully in
Sect.~\ref{sec:observing}.

Fig.~1 shows the distribution of transverse energy $E_T$ in the
control region cone
from the {\footnotesize HERWIG} soft underlying event (dashed curve)
and from the hard underlying event model (dotted curve).  The average
values, which are relevant for the correction to the inclusive jet
cross section as will be discussed in Sect.~\ref{sec:inclusive}, are
$\langle E_T \rangle = 0.84 \, {\rm GeV}$ for the soft background and
$\langle E_T \rangle = 1.90 \, {\rm GeV}$ for the hard background.
For comparison, Fig.~1 also shows the predicted contribution from
radiation associated with the hard scattering (solid curve), as given
by {\footnotesize HERWIG} with the underlying event turned off.
It has $\langle E_T \rangle = 2.37 \, {\rm GeV}$.

The hard underlying event model predicts a substantially larger
$\langle E_T \rangle$ than the {\footnotesize HERWIG} soft
model.  Taken as estimates of the average underlying event $E_T$ that
should be added to a perturbative QCD calculation to predict the
observed inclusive jet cross section, the difference is significant
as will be discussed in Sect.~\ref{sec:inclusive} --- the more so if
both mechanisms operate concurrently, so their contributions to $E_T$
should be added.

Before turning to the influence on inclusive jet cross sections, we
first consider how to measure the background event better.

\section {Observing the underlying event}
\label{sec:observing}

An obvious way to measure the background event contribution
to jets is to look at the $E_T$ distribution in a ``control region''
cone of the same radius that is used to identify jets, as discussed
in the preceeding Section.  Fig.~2 shows
the predicted probability distributions, which correspond to adding
the contributions described by Fig.~1:
{\footnotesize HERWIG} with its soft underlying event model included
(dashed curve),
{\footnotesize HERWIG} with its soft underlying event portion replaced
by the hard underlying event model of Sect.~\ref{sec:HUE}
(dotted curve), and
{\footnotesize HERWIG} with both underlying event models operative
(solid curve).
All three curves share contributions from the substantial QCD
radiation generated by the principal hard scattering and its color
connections to the beam particles.  This makes the curves somewhat
similar, but the differences are large enough that a measurement
of this control $E_T$ distribution would give a useful indication of
the background event level.

To measure the underlying event more accurately, we generalize a
technique first advocated by Marchesini and
Webber \cite{marchesini} --- which has apparently not yet been tried
experimentally in its original form.  The essence of the technique is
to study $E_T$ simultaneously in two regions that are separate from
each other and separate from any jets that define the final state
under study.

We describe the proposed generalization in the context of our
application to dijet events.  Let $(\eta_1,\phi_1)$ and
$(\eta_2,\phi_2)$ be the locations of the jets in the Lego plane.
For this study, we make the approximation that the jets are
back-to-back in azimuth:  $|\phi_1 - \phi_2| = \pi$.  Label the jets
so that $\eta_1 < \eta_2$ and let
\begin{eqnarray}
\kappa = \sqrt{4/[\pi^2 + (\eta_1 - \eta_2)^2] - 1/4} \;,
\label{eq:eq3}
\end{eqnarray}
rejecting the small fraction ($5.6 \%$) of events that have
$|\eta_1 - \eta_2| > 2.48$, for which $\kappa$ is undefined.  Then
consider the two points $(\eta_a,\phi_a)$ and $(\eta_b,\phi_b)$
where
\begin{eqnarray}
\eta_a &=& \eta_b = (\eta_1 + \eta_2)/2 + \kappa \pi \label{eq:eq4} \\
\phi_{a,b} &=& \phi_1 \pm (\pi/2 - \kappa |\eta_1 - \eta_2|) \;.
\label{eq:eq5}
\end{eqnarray}
These points are at the same pseudorapidity $\eta_a = \eta_b$, and
are well separated from each other in azimuth by
$|\phi_a - \phi_b| = \pi - 2 \kappa |\eta_1 - \eta_2|$ which varies
from $2.28$ to $3.14\,$.  Both points are separated from both of the
jet axes by a distance of exactly $2.0$ in the Lego plane.  (The jet
that lies farther from them in $\eta$ is therefore closer to them in
$\phi$.)  We can define a similar pair of points by letting
$\kappa \to -\kappa$ in Eq.~(\ref{eq:eq4}) and $\phi_1 \to \phi_2$
in Eq.~(\ref{eq:eq5}).  For the present purpose, we use only the pair
of points with $\eta_a = \eta_b$ closer to $\eta = 0$ to reduce any
kinematic suppression.

Let $S_a$ and $S_b$ denote the total transverse energy $E_T$ deposited
in cones of radius $0.7$ centered on these two points.  Fig.~3 shows
some typical locations of the cones with respect to the jet axes.
From the {\footnotesize HERWIG} simulation (taking the jet axes from
the directions of their partons, without modification by initial state
radiation for simplicity), we find $0 \pm 0.64$ for the mean and
standard deviation in $\eta$ and $2.59 \pm 0.26$ for the mean and
standard deviation in $|\phi_a - \phi_b|$.  (In practice, it may be
better to keep the control cones centered at fixed values of $\eta$,
at the expense of letting their distances from the jets vary somewhat,
to avoid effects due to the $\eta$-dependence of detector corrections.)

The cones $a$ and $b$ are those already used in Sect.~\ref{sec:HUE} to
study the underlying event background.  The two cones are equivalent,
and only one was used for each event.  Fig.~2 can therefore be
interpreted as the probability distribution for $S_a$ or equivalently
for $S_b$.

Because there is no intrinsic difference between the two control
cones, averaging over events would give
$\langle S_a \rangle = \langle S_b \rangle$.  It follows that
\begin{eqnarray}
\langle S_a \rangle = \langle S_b \rangle =
\left\langle \frac{S_a + S_b}{2} \right\rangle =
\left\langle S_{\rm diff} \right\rangle \, + \,
\left\langle S_{\rm min} \right\rangle
\label{eq:eq6}
\end{eqnarray}
where
\begin{eqnarray}
S_{\rm diff} &=& |S_a - S_b|/2 \label{eq:eq7}\\
S_{\rm min}  &=& \min(S_a,S_b) \;.
\label{eq:eq8}
\end{eqnarray}
This separation into $S_{\rm diff}$ and $S_{\rm min}$ is useful,
as pointed out in Ref.~\cite{marchesini}, because next-to-leading
order perturbative corrections to the principal hard scattering
contribute in at most one of the two regions, and hence contribute
only to $S_{\rm diff}$; while the underlying event (like minimum
bias events \cite{UA5corr}) is expected to have positive
correlations over long distances in $(\eta,\phi)$, so its
contribution to $S_{\rm diff}$ is suppressed while its contribution
to $S_{\rm min}$ is strong.  Thus measuring the distributions of
$S_{\rm diff}$ and $S_{\rm min}$ separately will be much more
revealing than the distribution of $S_a$ or $S_b$ of Fig.~2 alone.

Fig.~4 shows the predicted probability distributions for
$S_{\rm diff}$.  As anticipated above, $S_{\rm diff}$ is
not very sensitive to the choice of model for the underlying event.
Testing this distribution against experiment would therefore be a
good way to test QCD in a manner that is not very sensitive to
underlying event effects.

Fig.~5 shows predicted probability distributions for $S_{\rm min}$.
As anticipated above, $S_{\rm min}$ is {\it very} sensitive to the
underlying event model.  Testing this distribution against experiment
would therefore be an excellent way to measure the strength of the
underlying event.

The method of Marchesini and Webber \cite{marchesini} could be
generalized further in the hunt for the background.  For example,
the transverse energy could be measured in all four of the control
regions that can be defined by $R=0.7$ cones centered $2.0$ units in
$(\eta,\phi)$ from both jets.  These regions are shown in Fig.~3 for
jets that are back-to-back in $\phi$.  The distribution of the
minimum of the four transverse energies, or the distribution of the
sum of the two smaller ones, would be especially  sensitive to the
underlying event.  Other cone sizes
would also be of interest, but $R=0.7$ applies most directly to the
correction to the inclusive jet cross section.

\section {Correction to pQCD inclusive jet cross section}
\label{sec:inclusive}

The influence of the underlying event on the inclusive jet cross
section is calculated as follows.  Let
\begin{eqnarray}
F(E_T) \equiv \frac{d\,\sigma_{\rm single}}{d\,E_T}
\label{eq:eq9}
\end{eqnarray}
be the single hard scattering contribution to the jet cross
section, which is predicted by standard pQCD techniques on the basis
of parton distribution functions \cite{global}.  Let
\begin{eqnarray}
 G(E_T) \equiv \frac{d\,P}{d\,E_T}
\label{eq:eq10}
\end{eqnarray}
be the probability distribution for additional $E_T$ inside the jet
cone contributed by the underlying event, normalized to
$\int_0^\infty G(E) \, dE = 1$.  The observable jet $E_T$ is the total
of the two contributions, so the observable inclusive cross section
is given by
\begin{eqnarray}
F_{\rm obs}(E_T) \equiv
\frac{d\,\sigma_{\rm obs}}{d\,E_T} &=&
 \int_0^\infty dE_1 \, G(E_1) \,
 \int_0^\infty dE_2 \, F(E_2) \;
 \delta(E_1 + E_2 - E_T) \nonumber \\  &=&
\int_0^{E_T} G(E_1) \, F(E_T - E_1) \, dE_1 \;.
\label{eq:eq11}
\end{eqnarray}
It is assumed here that the underlying event contribution is small
enough not to significantly shift the apparent jet
axis.  In practice there would be a small additional increase in the
average jet energy due to the jet-finding algorithm's tendency to
maximize the $E_T$ included in the jet.

Fig.~6 shows the fractional increase in the inclusive jet cross
section caused by the {\footnotesize HERWIG} soft background event
(dashed curve);
by the hard underlying event model of Sect.~\ref{sec:HUE}
(dotted curve); or
by including both underlying event models
(dot-dash curve).
The effect of the underlying event is rather large at the lower values
of jet $E_T$, so it must be measured rather well before a meaningful
comparison can be made between the observed jet cross section and the
pQCD prediction.  The effect of the underlying event is significant
even at the highest jet $E_T$ shown, where the cross section is raised
by approximately $3\%$, $6\%$, or $9\%$ under these three assumptions.
The fractional increase caused by adding the hard underlying event to
the prediction with the soft underlying event already included is
shown by the solid curve in Fig.~6.  It is nearly identical to the
the increase caused by adding it to the pQCD prediction alone (dotted
curve).

The procedure used by experimenters \cite{cdf,d0} to take account
of the underlying event, along with background from simultaneous
events (``pile-up''), is to subtract a constant value from each
measured jet $E_T$.  We examine the accuracy of that approach next.

An average underlying event contribution $U$ can be defined by the
exact equation
\begin{eqnarray}
F_{\rm obs}(E_T) = F(E_T - U) \; .
\label{eq:eq12}
\end{eqnarray}
Since $G(E)$ falls rapidly with $E$, it is natural to make a linear
expansion of $F(E)$ in the neighborhood of $E_T$ in Eq.~(\ref{eq:eq11}).
The underlying event contribution is then given by
\begin{eqnarray}
U \cong \overline{E} = \int_0^\infty G(E) \, E \, dE \, ,
\label{eq:eq13}
\end{eqnarray}
which corresponds to the experimental procedure of approximating
the background contribution in each event by the average value.
This average is subtracted from each measured jet $E_T$ before
obtaining the inclusive cross section that is compared with pQCD.
The value used is $0.9 - 1.1 \, {\rm GeV}$, which is similar to
the {\footnotesize HERWIG} soft background prediction
$\langle E_T \rangle = 0.84 \, {\rm GeV}$ found above.
It corresponds to an underlying event level of
$\approx 0.6 - 0.7 \, {\rm GeV}$ per unit area in $(\eta,\phi)$.

I find that the linear approximation works rather well for the models
considered here, even in the most extreme case where both soft and
hard background are included.  For example, in that case
$\langle E_T \rangle = 0.84 + 1.90 = 2.74 \, {\rm GeV}$, while the
true effect corresponds to $U = 3.0$ for jets of
$E_T = 100 \, {\rm GeV}$.  However, in real experiments, there
is a further background from true minimum bias collisions that occur
between other $p \bar p$ pairs at high luminosity.  This can easily
raise the background level to the point where the linear
approximation breaks down.  In that case, the CDF analysis
method \cite{cdf} of parametrizing $F(E_T)$ and fitting the parameters
to the experimental results, which is useful to unfold other detector
effects anyway, can include this one as well.  The essential need is
to allow for the event-to-event fluctuations in background.

In the linear approximation, the fractional effect on the inclusive
cross section can be written in the form
\begin{eqnarray}
\frac{F_{\rm obs}(E_T) - F(E_T)}{F(E_T)}
\cong \frac{n \, U}{E_T}
\label{eq:eq14}
\end{eqnarray}
where $n$ is the local effective power law defined by
$F \propto E_T^{-n}$, i.e.
\begin{eqnarray}
n(E_T) = - d(\ln F)/d(\ln E_T) \; .
\label{eq:eq15}
\end{eqnarray}
Over the range $50 \, {\rm GeV} < E_T < 400 \, {\rm GeV}$, $n(E_T)$
rises from $\simeq \! 5.5$ to $\simeq \! 12.5$.  Its large value,
which represents the rapid fall of the inclusive cross section with
$E_T$, enhances the effect of the background according to
Eq.~(\ref{eq:eq14}), as has been emphasized recently \cite{soper}.
For example, it implies that an additional $1$--$2$ GeV of background
$E_T$, which may be present by the mechanism of Sect.~\ref{sec:HUE},
will raise the inclusive cross section by $3$--$6\%$ at
$E_T = 400 \, {\rm GeV}$, which is consistent with Fig.~6.

\section {Conclusions}
\label{sec:conclusions}

The point of this paper is that the inclusive jet cross section may
contain an underlying event contribution that is $1$--$2 \, {\rm GeV}$
larger than generally assumed, due to incoherent multiple semi-hard
interactions that accompany the hard scattering.  The calculation
presented here is less elaborate than previous methods of assessing
these multiparton interactions \cite{sjostrand,gyulassy,butterworth},
but it has the advantage of being constrained by a recent measurement
of the double parton interaction rate \cite{dpi}.

In view of the importance of the inclusive jet cross section, it is
urgent to settle the question of underlying event level definitively
by measuring it as described in Sect.~\ref{sec:observing}.  This will
also serve as a test of the assumptions used to make the estimate,
which involve interesting unexplored areas of non-perturbative QCD.

We have focused on $E_T$ in an $R = 0.7$ cone, because that is the
relevant quantity for estimating the incoherent background under
a jet.  The background will also have an important influence on
``jet shape'' observables such as the cone size dependence
$\psi(r) = \sum_{r_i < r} E^i_T / \sum_{r_i < R} E^i_T $ \cite{shape}.
To study the underlying event mechanism in more detail, it would be
useful to apply the technique based on Marchesini and
Webber \cite{marchesini} that is discussed in
Sect.~\ref{sec:observing} using regions of different sizes as well.
It would also be useful to apply the technique to different hard
processes such as $W$ and $W + \, 1 \, {\rm jet}$ production, and
for comparison to minimum bias events.

\section*{Acknowledgments}
I thank J. Huston and W.K. Tung for comments on the manuscript,
and H.L. Lai for supplying numerical values for the
CTEQ4M inclusive jet prediction.
This work was supported in part by U.S. National Science Foundation
grant number PHY--9507683.

\newpage

\begin{figure}
    \begin{center}
       \leavevmode
       \epsfxsize=1.0\hsize
      \epsfbox{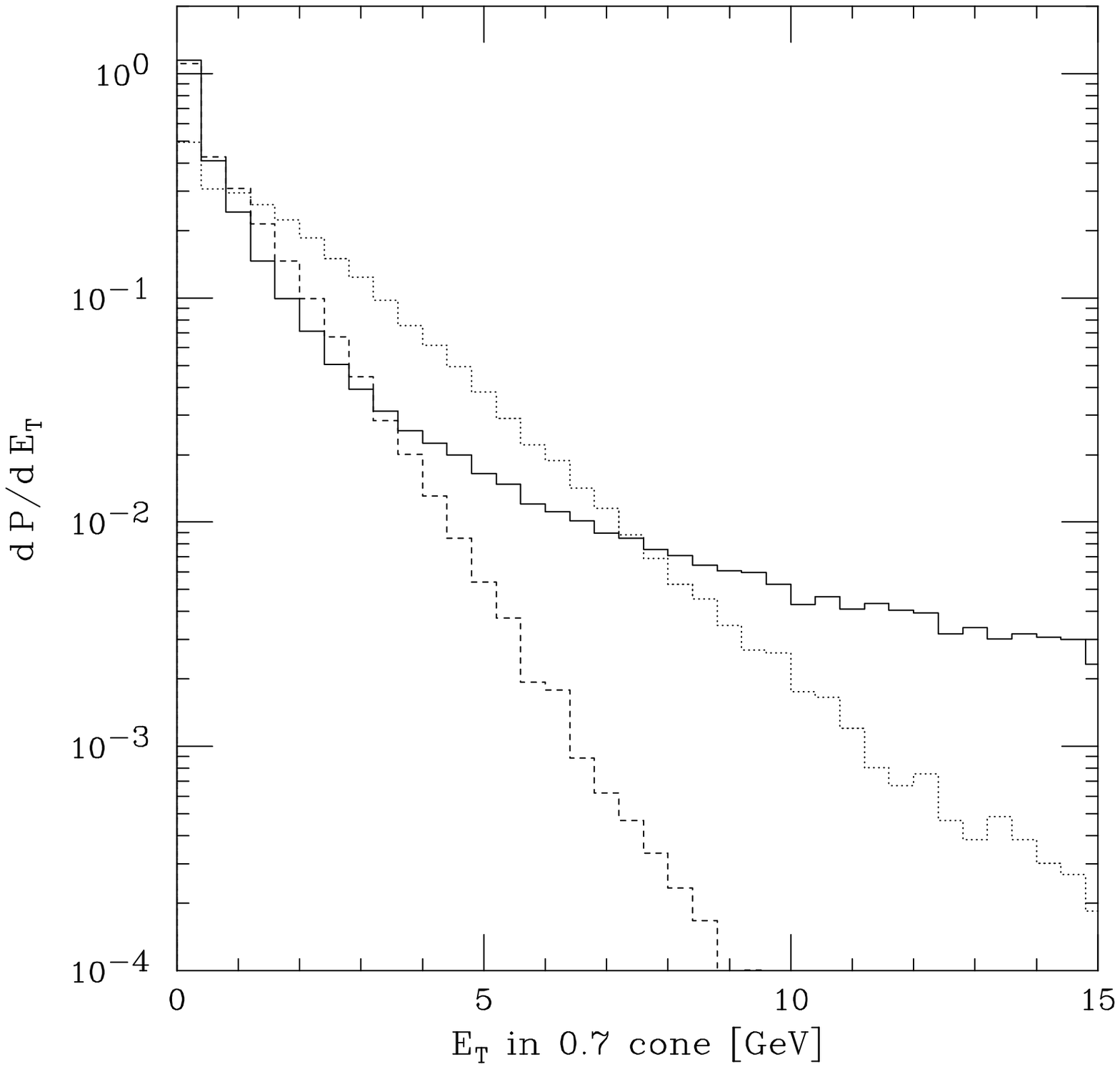}
    \end{center}
\caption{
Normalized probability distributions for $E_T$ in $R=0.7$
control cone.
Solid curve = {\protect\footnotesize HERWIG} with soft underlying
event off;
dashed curve =  soft underlying event from
{\protect \footnotesize HERWIG};
dotted curve =  hard underlying event from
Sect.~{\protect \ref{sec:HUE}}.
}
\label{figure1}
\end{figure}

\begin{figure}
    \begin{center}
       \leavevmode
       \epsfxsize=1.0\hsize
      \epsfbox{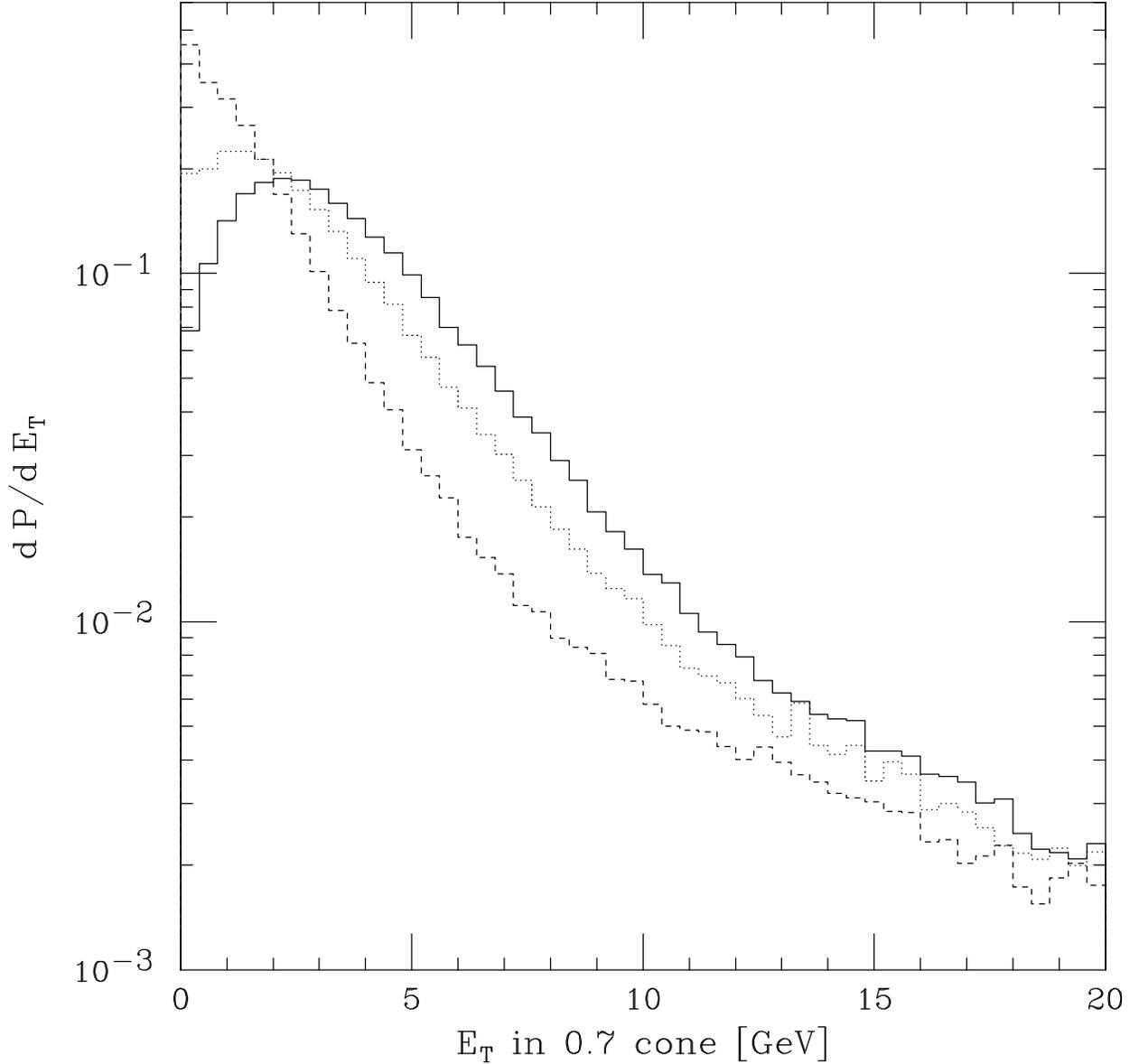}
   \end{center}
\caption{
Predicted probability distributions for $E_T$ in $R=0.7$
control cone.
Dashed curve = {\protect \footnotesize HERWIG} including
soft underlying event;
dotted curve =  {\protect \footnotesize HERWIG} with soft
underlying event replaced by hard underlying event model of
Sect.~{\protect \ref{sec:HUE}};
solid curve = {\protect \footnotesize HERWIG} with both soft and
hard underlying event included.
}
\label{figure2}
\end{figure}

\begin{figure}
    \begin{center}
    \begin{tabular}{ccc}
       \mbox{
          \leavevmode
          \epsfxsize=0.3\hsize
          \epsfbox{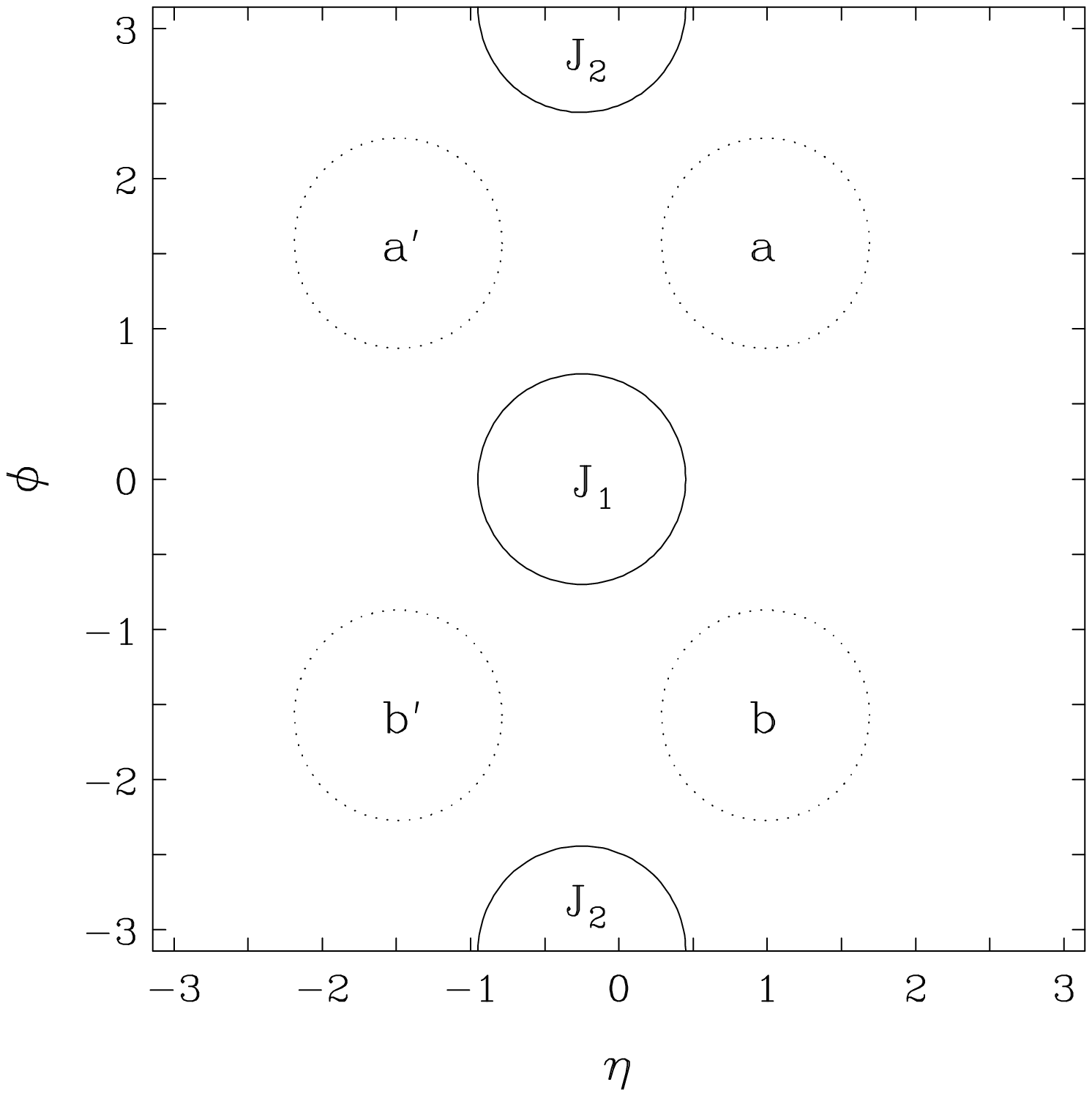}
            } &
       \mbox{
          \leavevmode
          \epsfxsize=0.3\hsize
          \epsfbox{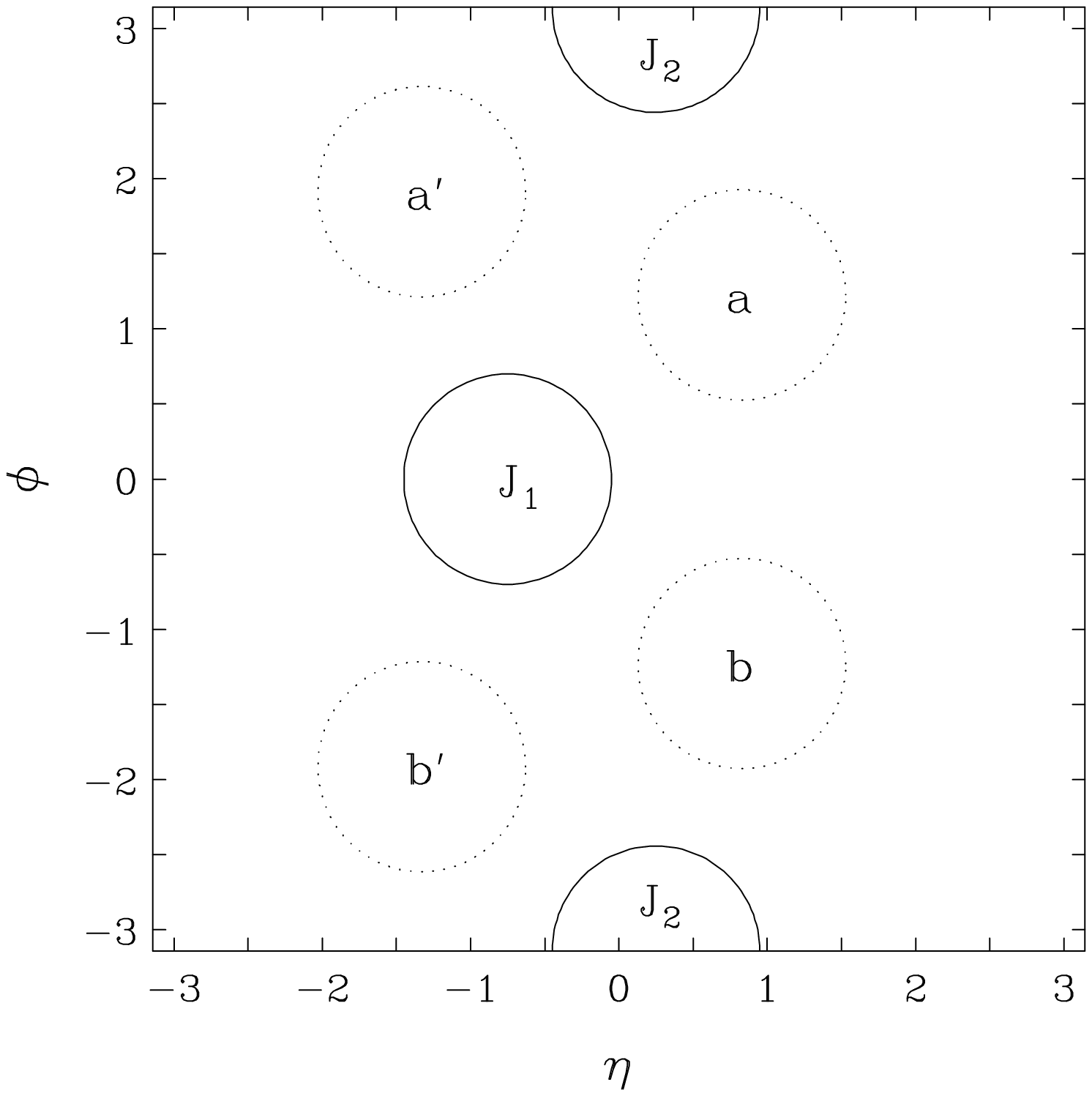}
            } &
        \mbox{
          \leavevmode
          \epsfxsize=0.3\hsize
          \epsfbox{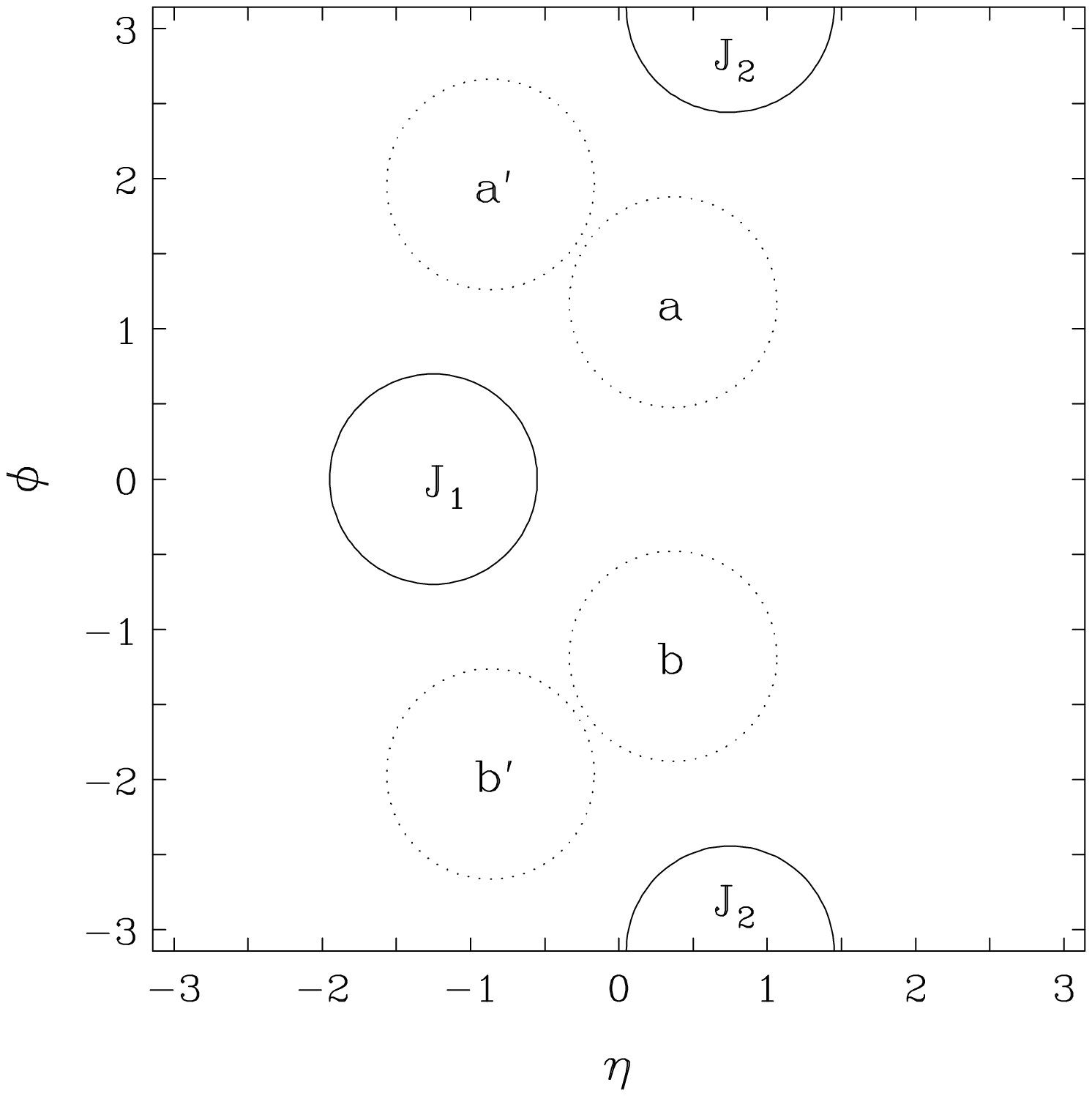}
             }
    \end{tabular}
    \end{center}
\caption{
Typical ``control'' regions $a, b, a^\prime, b^\prime$ for measuring
the transverse energies $S_a$ and $S_b$.  Configurations shown are for
jets $J_1$ and $J_2$ with $|\eta_1 - \eta_2| = 0.0, 1.0, 2.0$.
The centers of the control regions are a distance $2.0$ from both jet
axes.  The distance from $a$ to $b$ is the same as the distance from
$a^\prime$ to $b^\prime$, although this is not apparent because the
azimuthal cylinder is cut for display at $|\phi| = \pi$.  We use the
pair $ab$ or $a^\prime b^\prime$ (here $ab$) that is closer to
$\eta = 0$.
}
\label{figure3}
\end{figure}

\newpage

\begin{figure}
    \begin{center}
       \leavevmode
       \epsfxsize=1.0\hsize
      \epsfbox{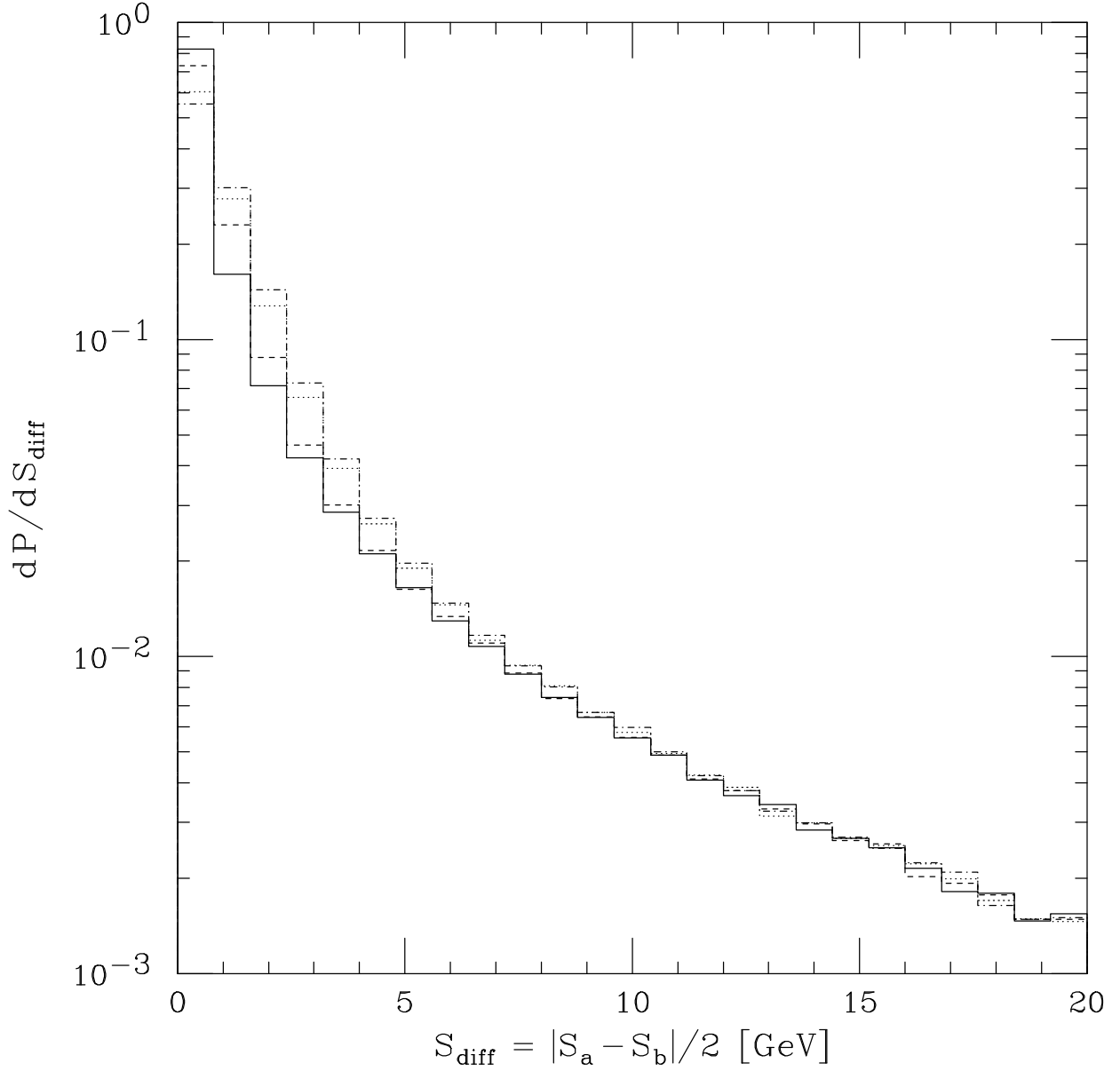}
    \end{center}
\caption{
Normalized probability distributions for
$S_{\rm diff} = |S_a - S_b|/2$,
where $S_a$ and $S_b$ are the total transverse energies in the
two $R=0.7$ control regions.
Solid curve = {\protect\footnotesize HERWIG} with soft
underlying event off;
dashed curve = {\protect\footnotesize HERWIG} with soft
underlying event on;
dotted curve = {\protect\footnotesize HERWIG} with soft
underlying event replaced by hard underlying event from
Sect.~{\protect \ref{sec:HUE}};
dot-dash curve = {\protect\footnotesize HERWIG} with
both soft and hard underlying event.
}
\label{figure4}
\end{figure}

\begin{figure}
    \begin{center}
       \leavevmode
       \epsfxsize=1.0\hsize
      \epsfbox{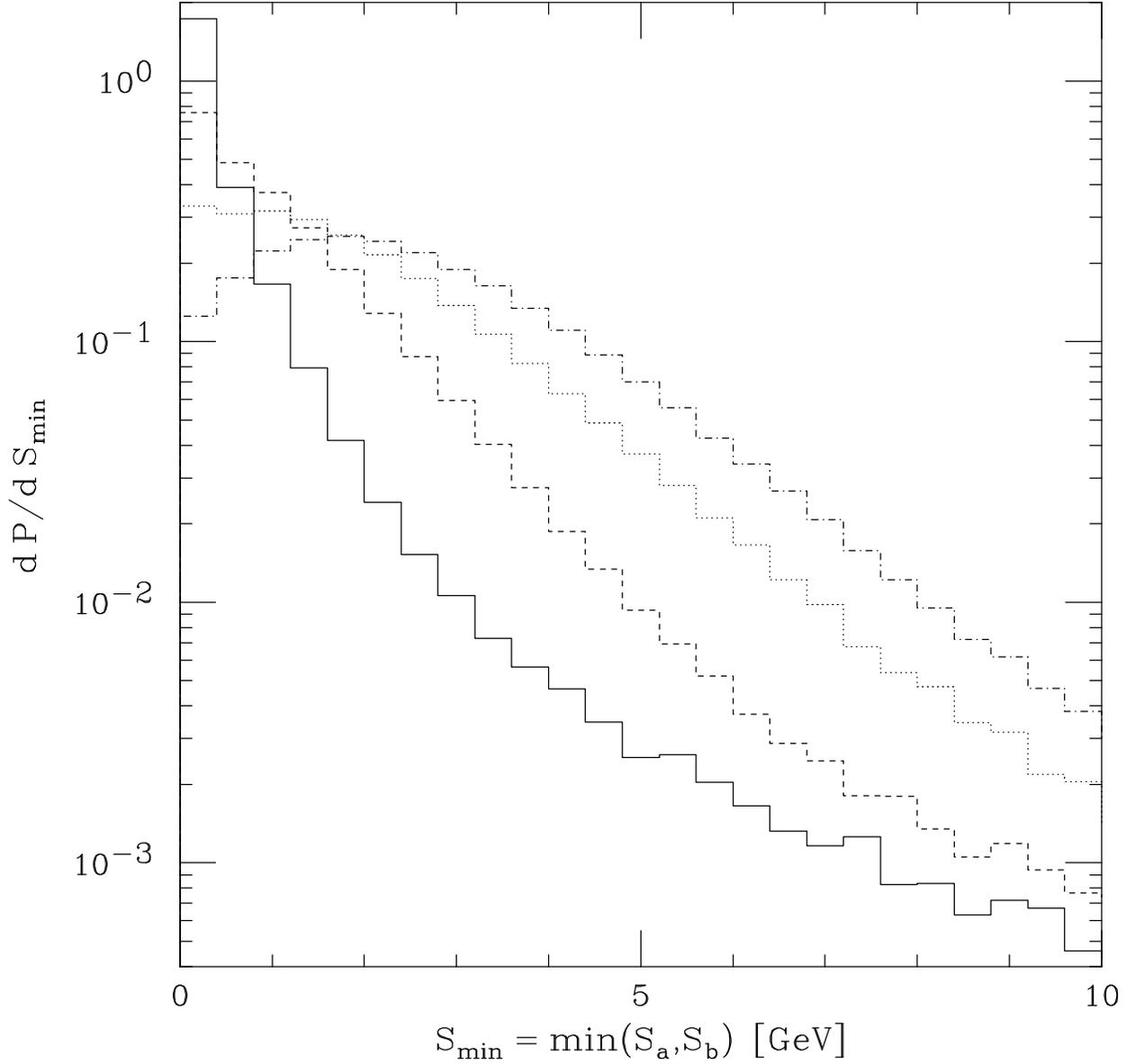}
    \end{center}
\caption{
Normalized probability distributions for
$S_{\rm min} = \min(S_a,S_b)$.
As in Fig.~4,
Solid curve = {\protect\footnotesize HERWIG} with soft
underlying event off;
dashed curve = {\protect\footnotesize HERWIG} with soft
underlying event on;
dotted curve = {\protect\footnotesize HERWIG} with soft
underlying event replaced by hard underlying event from
Sect.~{\protect \ref{sec:HUE}};
dot-dash curve = {\protect\footnotesize HERWIG} with
both soft and hard underlying event.
}
\label{figure5}
\end{figure}

\begin{figure}
    \begin{center}
       \leavevmode
       \epsfxsize=1.0\hsize
      \epsfbox{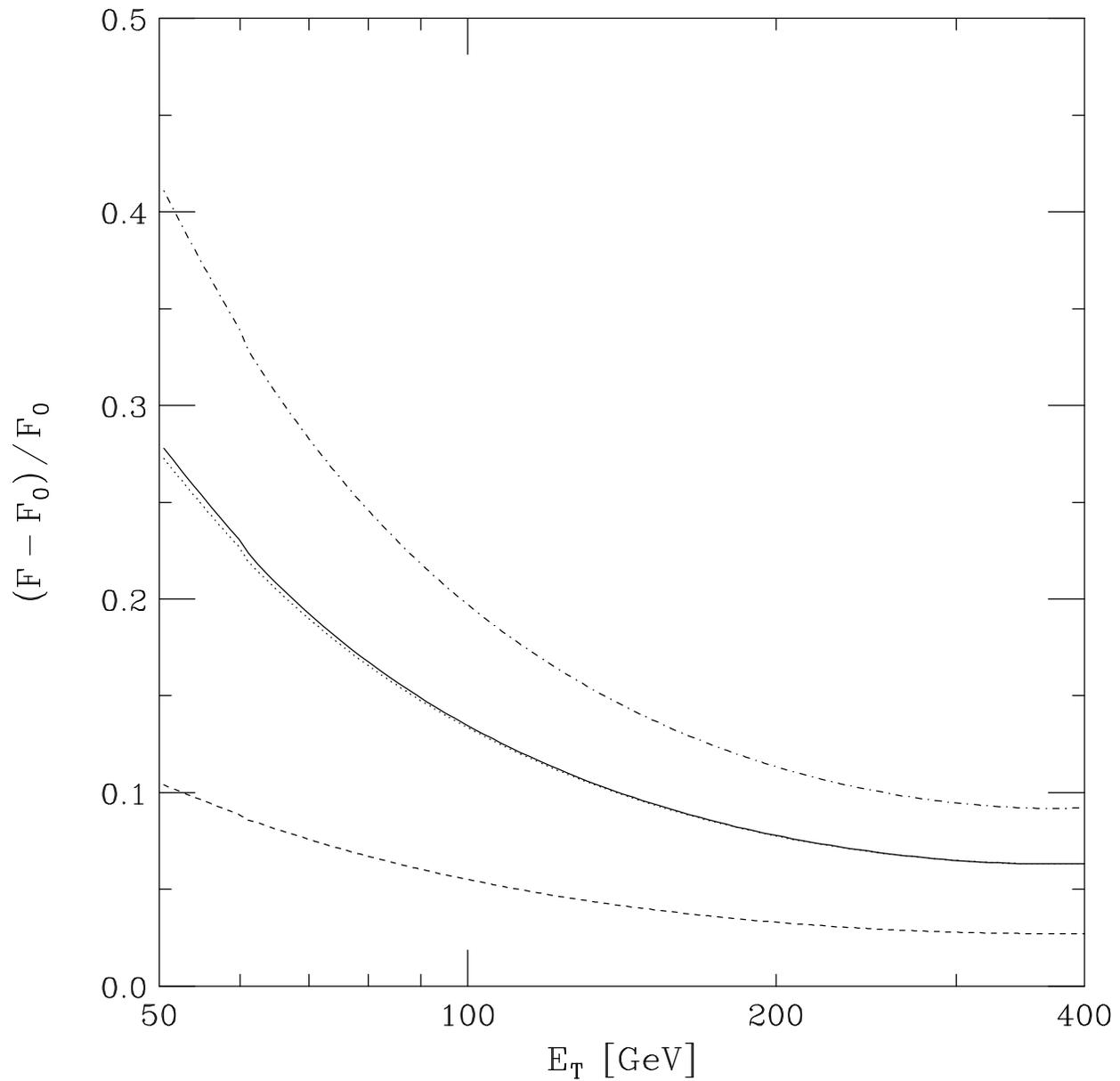}
    \end{center}
\caption{
Fractional increase in the inclusive jet cross section produced by
adding {\protect\footnotesize HERWIG} soft underlying event
(dashed curve);
the hard underlying event model of
Sect.~{\protect \ref{sec:HUE}} (dotted curve); or
both underlying event contributions (dot-dash curve) to the
pQCD prediction.  The fractional increase produced by adding
the hard underlying event to the soft+pQCD prediction is shown
by the solid curve.
}
\label{figure6}
\end{figure}

\end{document}